\documentclass[journal]{IEEEtran}
\usepackage{graphicx}
\usepackage{caption}
\usepackage{subcaption}
\usepackage{booktabs}
\usepackage{tikz}
\usepackage{xcolor}

\newcommand{\circled}[1]{\tikz[baseline=(myanchor.base)] \node[circle,fill=.,inner sep=1pt] (myanchor) {\color{-.}\bfseries\footnotesize #1};}

\usepackage{cite}

\ifCLASSINFOpdf
\else
\fi

\begin{document}
%
\title{Fixed-Posit: A Floating-Point Representation for Error-Resilient Applications}
%
%
%

\author{Varun~Gohil*\thanks{*Authors contributed equally}, Sumit~Walia*, Joycee~Mekie, Manu~Awasthi\vspace{-20pt}
\thanks{This work is supported through grants received from SMDP-C2SD and  YFRF under Visvesvaraya PhD scheme, Ministry of Electronics and IT (MEITY), through SERB grants CRG/2018/00501, MTR/2019/001605 at the Indian Institute of Technology, Gandhinagar, Ashoka University startup grant and joint-grant received from the Semiconductor Research Corporation (SRC), through contracts 2020-IR-2980 with the Indian Institute of Technology, Gandhinagar, and 2020-IR-3005 with Ashoka University.}
\thanks{V. Gohil and M. Awasthi are with Department of Computer Science, Ashoka University, Haryana, India, 131029. (e-mail: \{varun.gohil, manu.awasthi\}@ashoka.edu.in). 
Part of this work was done when V. Gohil was a student at IIT Gandhinagar. 
S. Walia and J. Mekie are with Department of Electrical Engineering, Indian Institute of Technology Gandhinagar, Gujarat, India, 382355. (e-mail: \{sumit.walia, joycee\}@iitgn.ac.in).}
\thanks{© 2021 IEEE.  Personal use of this material is permitted.  Permission from IEEE must be obtained for all other uses, in any current or future media, including reprinting/republishing this material for advertising or promotional purposes, creating new collective works, for resale or redistribution to servers or lists, or reuse of any copyrighted component of this work in other works.}
}

\maketitle
\begin{abstract}
Today, almost all computer systems use IEEE-754 floating point to represent real numbers. Recently, posit was proposed as an alternative to IEEE-754 floating point as it has better accuracy and a larger dynamic range. The configurable nature of posit, with varying number of regime and exponent bits, has acted as a deterrent to its adoption. To overcome this shortcoming, we propose \textit{fixed-posit} representation where the number of regime and exponent bits are fixed, and present the design of a fixed-posit multiplier. We evaluate the fixed-posit multiplier on error-resilient applications of AxBench and OpenBLAS benchmarks as well as neural networks. The proposed fixed-posit multiplier has 47\%, 38.5\%, 22\% savings for power, area and delay respectively when compared to posit multipliers and up to  70\%, 66\%, 26\% savings in power, area and delay respectively when compared to 32-bit IEEE-754 multiplier. These savings are accompanied with minimal output quality loss (1.2\% average relative error) across OpenBLAS and AxBench workloads. Further, for neural networks like ResNet-18 on ImageNet we observe a negligible accuracy loss (0.12\%) on using the fixed-posit multiplier.
\end{abstract}


%
\IEEEpeerreviewmaketitle

\section{Introduction}
%
%
%
%
\IEEEPARstart{C}{omputers} cannot represent all of the infinitely many real numbers because they use a finite number of bits. Generally, computers represent a finite set of real numbers by creating a mapping between these numbers and the possible bit permutations. We use the term \textit{representation} to refer to this mapping. The chosen representation impacts the entire computing stack - from circuits which perform arithmetic, all the way to data types in programming languages. As a result, the choice of representation is critical to the design of efficient software and hardware.

Today, virtually all computer systems represent real numbers using IEEE-754 floating point representation\cite{ieee754}. Despite its popularity, it has many shortcomings. It breaks the mathematical laws of commutativity and associativity~\cite{goldberg} and  suffers from underflow / overflow while performing arithmetic operations\cite{goldberg, mata}. Further, since it has a fixed number of precision bits, it suffers from rounding errors when representing real numbers~\cite{goldberg}. Finally, IEEE-754 floating point reserves multiple bit patterns to represent Not-a-Number (NaN), and positive and negative zeros~\cite{goldberg}. This leads to wastage of bit-patterns, which could have been used to represent other numbers. Owing to these shortcomings, the hardware needs to handle rounding, NaNs, exceptions and many other corner-cases. Overall, this makes the design and verification of the IEEE-754 Floating Point Unit (FPU) a time-consuming and complex task.

The drawbacks of IEEE-754 have led to increasingly complex hardware designs. This, in part has resulted in the FPU being a major contributor to the processor's energy and area consumption~\cite{farhad}. Single precision FPU occupies 30-40\% of the die area, which increases to 50-55\% for a double precision FPU~\cite{farhad}. For embedded applications, floating point computations constitute 50\% of core and data memory's energy consumption \cite{energy}.

Gustafson et. al. \cite{gustafson} proposed \textit{posit} as a replacement for IEEE-754 floating point. Compared to IEEE-754 floating point, posits have larger dynamic range, higher accuracy and follow mathematical laws of commutativity and associativity\cite{gustafson,farhad}. Posits do not suffer from underflow or overflow while performing arithmetic operations\cite{gustafson}. Posits only have a single Not-a-Real (NaR) exception and only a single representation for zero, unlike IEEE-754 floating point\cite{gustafson}. This reduces the number of corner cases the hardware must handle, thereby making hardware design simpler.

Several recent works \cite{ml1,ml3,ml4, ml5} have used posit in hardware accelerators for deep learning workloads. However, adoption of posit  for processors running general purpose workloads has been a topic of debate owing to tradeoffs involved. The configurable bit format of posit - having varying number of regime, exponent and fraction bits, has acted as a deterrent to it's adoption in general-purpose processors. The configurable bit format requires sequential bit decoding, which adds to the critical path delay of the FPU. It further requires hardware to support extreme configurations which worsens its area and power consumption.

In this paper, we propose fixing the number of regime, exponent and fraction bits of posit to overcome its shortcomings. We refer to this as \textit{fixed-posit} representation. The major contributions of this work are: 
\begin{itemize}
    \item We find multiple fixed-posits having the same exponent range as that of IEEE-754 floating point (-126 -- 127), which makes them candidates to replace IEEE-754 floating point.
    \item We show that configurable nature of posits leads to posit multipliers having worse delay, area and power compared to 32-bit IEEE-754 floating point multipliers. 
    \item We show that fixed-posit results in more efficient multipliers than posit, in terms of power, area and delay.
    \item We compare fixed-posit multipliers of varying bit-widths with 32-bit IEEE-754 floating point multiplier using OpenBLAS and AxBench benchmarks. We show that fixed-posit multipliers can lead to up to  70\%, 66\%, 26\% savings in power, area and delay respectively with minimal quality loss (1.2\% average relative error).
    \item We show the efficacy of fixed-posits on machine learning applications. Using fixed-posits for a 3 layered fully-connected network performing inference on MNIST leads to 0.04\% accuracy drop. We observe negligible accuracy loss (~0.12\%) even for larger models like ResNet-18 on ImageNet.
\end{itemize}

\begin{figure}
    \centering
    \subfloat[IEEE-754 representation\label{fig:ieeebit}]{%
       \includegraphics[width=0.6\linewidth]{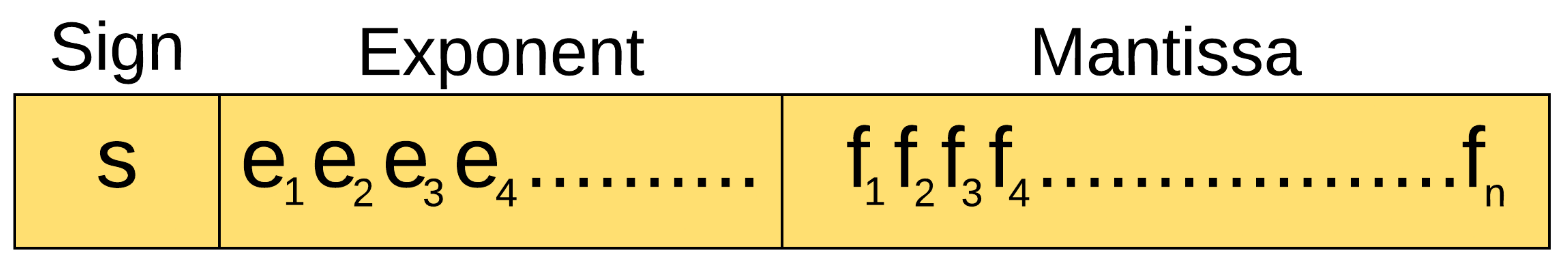}
       }
       \\
    \subfloat[Posit representation\label{fig:positbit}]{%
       \includegraphics[width=0.6\linewidth]{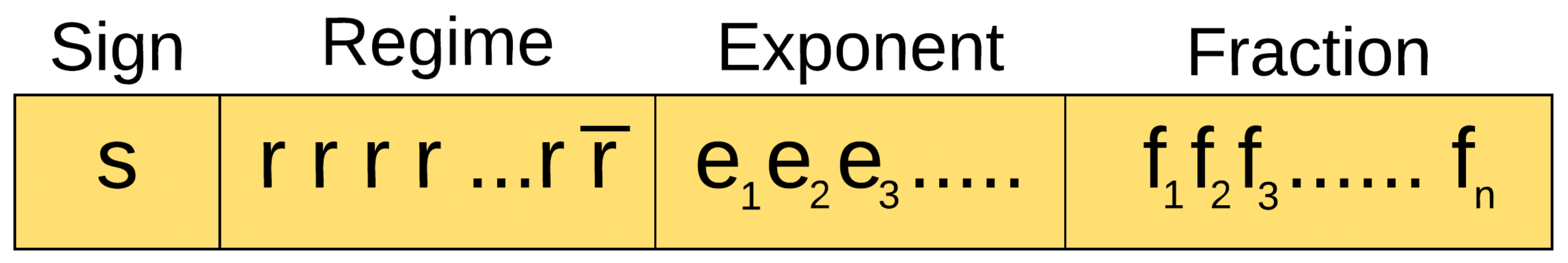}
       }
       \\
    \subfloat[Fixed-posit representation\label{fig:fixedpositbit}]{%
       \includegraphics[width=0.6\linewidth]{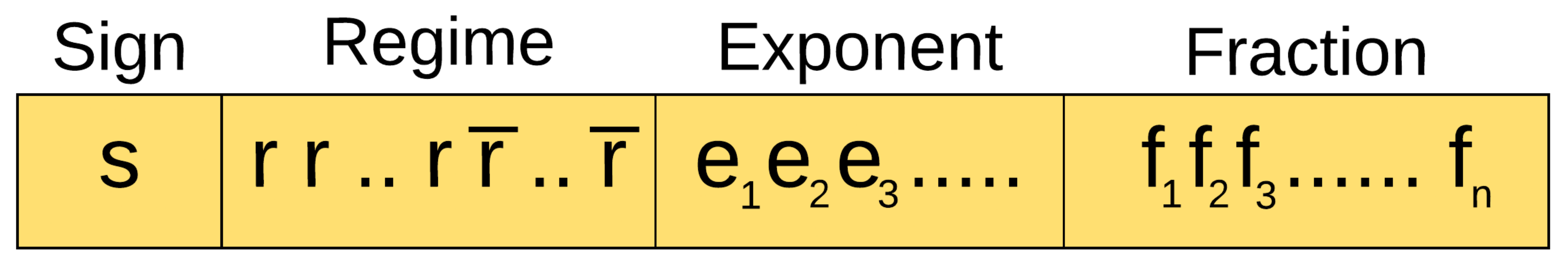}
       }
       \caption{Bit Formats}
      \vspace{-0.5cm}
\end{figure}

\section{Background}
\label{sec:bgposit}
Figure \ref{fig:ieeebit} shows that the 3 parts of IEEE-754 floating point representation: sign, exponent and mantissa. Single precision floats have 1 sign bit, 8 exponent bits and 23 mantissa bits. For more details we refer the reader to \cite{ieee754,goldberg}.  

The posit representation is   represented using a 2-tuple $(N,es)$, where $N$ is the total number of bits and $es$ is the maximum number of exponent bits. It is divided into 4 parts - sign, regime, exponent and fraction, as shown in Figure \ref{fig:positbit}. The sign bit (S) is zero for non-negative numbers and 1 otherwise. The number of regime bits are variable and have a specific encoding. The regime is sequence of 0 or 1 bits and ends with the opposite bit. After regime, if there are at least \textit{es} bits remaining, the next \textit{es} bits belong to the exponent. If fewer than \textit{es} bits are left, the bits which remain belong to the exponent. The bits remaining after the exponent are fraction bits. The value \textit{x} represented in posit format is given by, 
$
    \textit{x} = (-1)^S \times (2^{2^{es}})^k  \times 2^{exponent} \times \left( 1 + \sum_{i = 1}^{n}f_{n-i}2^{-i} \right),
$
 where the value of \textit{k} is decided by the length of the regime. Let \textit{m} be the number of consecutive 0's or 1's bits in the regime; if regime bits are 0, then $k = -m$ and if regime bits are 1, then $k = m-1$\cite{gustafson}.  

\section{Fixed-Posit Representation}
\label{sec:mul}
The configurable bit format of posits, having varying number of regime, exponent
and fraction bits,  has acted as a deterrent to adoption of posits over floats,
for multiple reasons. We demonstrate this by comparing 32-bit IEEE-754 with
(32,6) posit. We choose (32,6) posit for comparison since it is optimal in terms
of delay, area and power as shown in Table \ref{tab:regime}. We do not consider
posits with  \textit{es} values greater than 6, like (32,7) since these posits
can have maximum 22 fraction bits thereby resulting in non-zero relative error.
First, the configurable format of posits requires sequential decoding of the bits, unlike IEEE-754 floating point where bits are decoded in parallel. Such sequential decoding of bits increases critical path delay of the FPU. Table~\ref{tab:regime} shows that critical path delay of 32-bit IEEE-754 floating point multiplier is 61\% lower than that of (32, 6) posit. 

Second, the hardware for posit needs to be designed to handle extreme configurations which results in significant penalties in area and power consumption.  For example, for a (32, 6) posit, hardware needs to handle one extreme configuration with 23 fraction bits as well as the other extreme configuration with 31 regime bits. To support such configurability, the hardware needs to have a 23-bit multiplier for fraction bits and 31-bit decoder for regime
bits. Table~\ref{tab:regime} shows that such a design for (32, 6) posit
multiplier consumes 94\% more power and 78\% more area than 32-bit IEEE-754
floating point multiplier. Further, it might be the case that the numbers
represented by the extreme configurations of posit lie outside the dynamic range
of IEEE-754 floating point and might not occur in any application developed using
IEEE-754 floating point~\cite{gustafson}. In such cases, supporting extreme
configurations would result in inefficient utilization of hardware and
unnecessarily add overhead in hardware design process. Overall, the results demonstrate that from a hardware design perspective, posit is not a viable replacement for IEEE-754 floating point in general-purpose processors.

 To overcome the shortcomings of posit, we fix the number of regime and exponent bits in the posit representation. This allows us to implicitly fix the number of fractions bits. We refer to such a representation as \textit{fixed-posit} representation. The regime in fixed-posits is a sequence of 0 or 1 bits similar to posits. However, as shown in Figure \ref{fig:fixedpositbit}, if sequence length is shorter than the fixed number of regime bits, the remaining positions are filled with complement bits (1's if regime sequence has 0's and vice-versa). One can decode a fixed-posit by the same decoding formula used for posits.  We represent a fixed-posit with a 3-tuple (Bit-width, exponent bits, regime bits).

\begin{figure}[]
    \centering
  \includegraphics[width=0.5\linewidth]{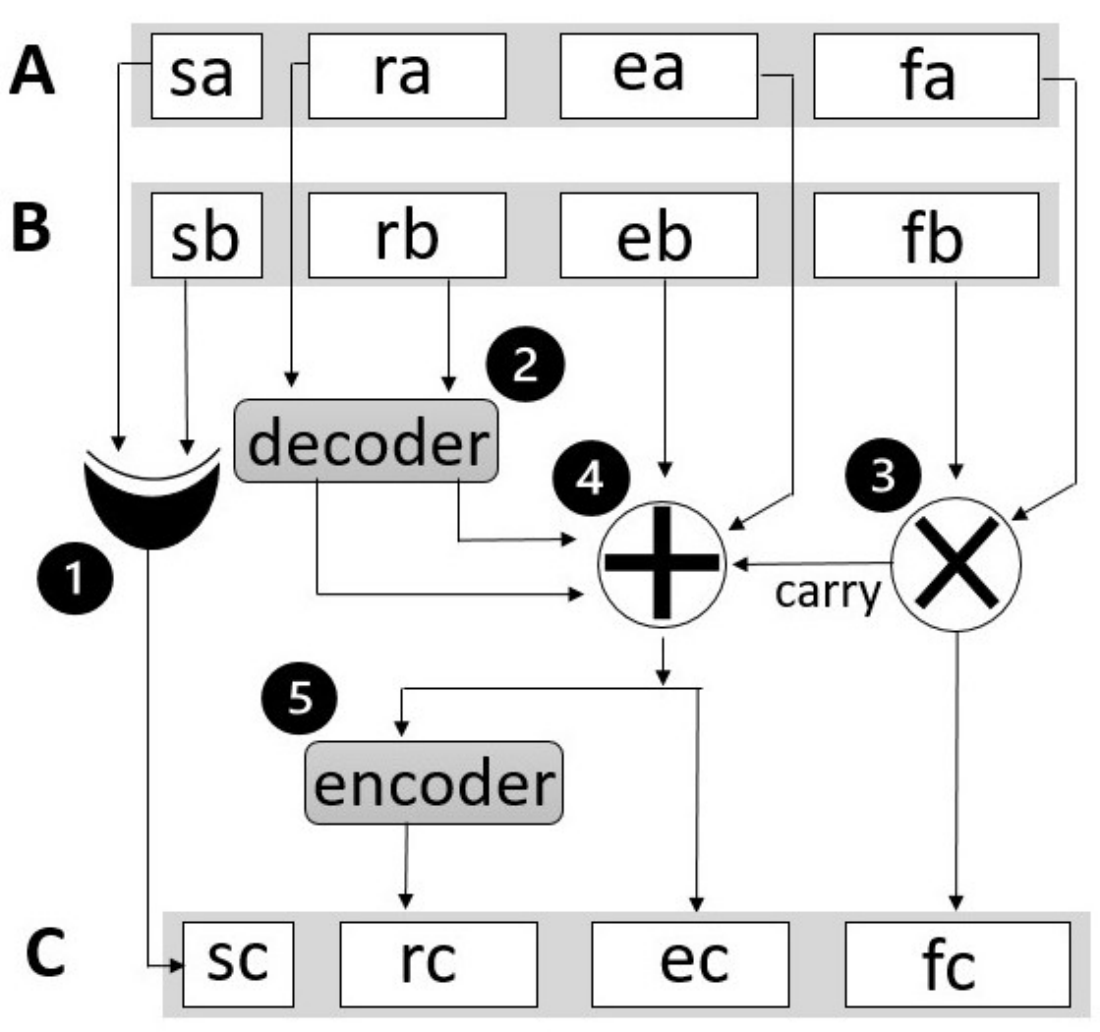}

  \caption{\centering Circuit diagram of fixed-posit multiplier}
  \label{fig:circuit}
  \vspace{-0.5cm}
\end{figure}

Figure \ref{fig:circuit} shows the block diagram of our fixed-posit multiplier. The diagram assumes that two operands \textit{A} and \textit{B} are multiplied to obtain the result \textit{C}. We use prefixes to indicate various part of the fixed-posit. Prefix \textit{s} refers to the sign bit, \textit{r} refers to regime, \textit{e} refers to exponent and \textit{f} refers to fraction.

As shown in Figure \ref{fig:circuit}, we obtain the sign bit of result (sc) by applying XOR operation on the sign bits of operands \textit{A} and \textit{B} \circled{1}. The decoder \circled{2} extracts the k-values of both operands and performs left-shift by the number of exponent bits. Next, the fraction bits of the operands (fa and fb) are multiplied and normalized \circled{3} to obtain the fraction bits of the result (fc) and the carry. Finally, the exponents of both operands, their left-shifted k-values and carry are added \circled{4}. The adder output has two components - the exponent of the result (ec) and the shifted-k value of the result. Finally, the encoder \circled{5} encodes the shifted k-value of result to obtain the result's regime bits (rc). Overall, our fixed-posit multiplier design is very similar to that of a IEEE-754 floating point multiplier. The major difference is the existence of encoder and decoder to handle the regime bits. 

The fixed-posit multiplier has simpler implementation as we can use shift registers in decoding regime bits, in place of leading zero and one detector used in posit multipliers\cite{peri}. Further, for fixed-posits, we use a smaller multiplier for the fraction bits in comparison with variable size of multiplier for posits. This results in the fixed-posit multiplier being smaller and faster than posit multiplier.

\vspace{-0.3cm}
\section{Evaluation of Fixed-Posits}

\begin{table}[t]
\Large
\centering
\caption{\centering Fixed-posits with same exponent range as 32-bit IEEE-754 floating-point (-126 to +127)}
\label{tab:list}
\resizebox{1\linewidth}{!}{%
\begin{tabular}{@{}ccclllccll@{}}
\toprule
\multicolumn{10}{c}{\textbf{Fixed-Posit : (Bit-width, exponent bits, regime bits)}}                                                                                        \\ \midrule
(32, 3, 16) & (32, 4, 8) & (32, 5, 4) & \multicolumn{1}{c}{(32, 6, 2)} & \multicolumn{1}{c|}{(32, 7, 1)} & (24, 3, 16) & (24, 4, 8) & (24, 5, 4) & (24, 6, 2) & (24, 7, 1) \\
(30, 3, 16) &
  (30, 4, 8) &
  (30, 5, 4) &
  (30, 6, 2) &
  \multicolumn{1}{c|}{(30, 7, 1)} &
  \multicolumn{1}{c}{(22, 3, 16)} &
  (22, 4, 8) &
  (22, 5, 4) &
  \multicolumn{1}{c}{(22, 6, 2)} &
  (22, 7, 1) \\
(28, 3, 16) & (28, 4, 8) & (28, 5, 4) & \multicolumn{1}{c}{(28, 6, 2)} & \multicolumn{1}{l|}{(28, 7, 1)} &             & (20, 4, 8) & (20, 5, 4) & (20, 6, 2) & (20, 7, 1) \\
(26, 3, 16) &
  (26, 4, 8) &
  (26, 5, 4) &
  (26, 6, 2) &
  \multicolumn{1}{l|}{(26, 7, 1)} &
  \multicolumn{1}{c}{} &
  (18, 4, 8) &
  \multicolumn{1}{l}{(18, 5, 4)} &
  (18, 6, 2) &
  (18, 7, 1) \\ \bottomrule
\end{tabular}%
}
\end{table}

\begin{table}[t]
\Large
\centering
\caption{\centering Comparison of representations (V: Variable)}
\label{tab:regime}
\resizebox{1\linewidth}{!}{%
\begin{tabular}{@{}ccccccccc@{}}
\toprule
\textbf{\begin{tabular}[c]{@{}c@{}}Type\end{tabular}} &
  \textbf{\begin{tabular}[c]{@{}c@{}}Bit-width\\ (N)\end{tabular}} &
  \textbf{\begin{tabular}[c]{@{}c@{}}Exponent\\ Bits\end{tabular}} &
  \textbf{\begin{tabular}[c]{@{}c@{}}Regime\\ Bits\end{tabular}} &
  \textbf{\begin{tabular}[c]{@{}c@{}}Fraction\\ Bits\end{tabular}} &
  \textbf{\begin{tabular}[c]{@{}c@{}}Relative \\ Error\end{tabular}} &
  \textbf{\begin{tabular}[c]{@{}c@{}}Power\\ ($\mu W$)\end{tabular}} &
  \textbf{\begin{tabular}[c]{@{}c@{}}Area\\ ($\mu m^2$)\end{tabular}} &
  \textbf{\begin{tabular}[c]{@{}c@{}}Delay\\ (ns)\end{tabular}} \\ \midrule
IEEE-754 & 32 & 8 & - & 23 & 0\%                       & 129.37 & 2923 & 0.54 \\ \midrule
  & 32 & $\leq 2$ & V & V & 0\%                       & 319.82 & 6198 & 0.93 \\
  & 32 & $\leq 3$ & V & V & 0\%                       & 294.9 & 5672 & 0.90 \\
Posit& 32 & $\leq 4$ & V & V & 0\%                       & 284.6 & 5384 & 0.88 \\
 & 32 & $\leq 5$ & V & V & 0\%                       & 261.4 & 5367 & 0.87 \\
 & 32 & $\leq 6$ & V & V & 0\%                       & 250.8 & 5200 & 0.87 \\ \midrule
   & 32 & 3        & 16       & 12       & $2.44 \times 10^{-2}\%$ & 109.83 & 2312 & 0.52 \\
  & 32 & 4        & 8        & 19       & $1.78 \times 10^{-4}\%$ & 117.45 & 2834 & 0.56 \\
Fixed-        & 32 & 5        & 4        & 22       & $1.19 \times 10^{-5}\%$ & 134.97  & 3301 & 0.58 \\
Posit        & 32 & 6        & 2        & 23       & $0$\% & 132.80  & 3196 & 0.56 \\
    & 32 & 7        & 1        & 23       & $0$\% & 135.32  & 3298 & 0.57 \\ \bottomrule
\end{tabular}%
}
 \vspace{-0.5cm}
\end{table}

To study fixed-posit as a potential replacement of IEEE-754 floating point, we find multiple fixed-posit representations having the same exponent range as 32-bit IEEE-754 floating point (-126 to +127). In fixed-posits, the regime would also contribute to the exponent range. We present the list of such fixed-posit representations in Table \ref{tab:list}. We use this list of fixed-posits for further experiments.
\vspace{-0.4cm}
\subsection{Comparison with Posits}
\label{sec:compposit}
We compare the multipliers of fixed-posits in  Table \ref{tab:list} with multiplier of posits based on their power, area and delay. We further compute the relative error when multiplication operands are converted from 32-bit IEEE-754 floating point to fixed-posit/conventional posit. The relative error between a number \textit{x} in IEEE-754 and its representation \textit{x'} in fixed-posit/conventional posit is $100 \times |\textit{x} - \textit{x'}|/|\textit{x}|$.

For our experiments, we design the multipliers using Verilog. We synthesize them on 28nm FDSOI technology node at 1.2V supply voltage and at $25^o$C using the Synopsys Design Compiler. We compute the total power by testing the multiplier on 100K single precision IEEE-754 floating point numbers generated uniformly at random within the range $2^{-126}-2^{127}$.

Table \ref{tab:regime} shows the results of our analysis for 32-bit fixed-posit. As results show, more number of fraction bits offer more precision thereby leading to lower errors. However, one needs a larger multiplier for multiplying more number of fraction bits that also results in increased power consumption. On the other hand, fixed-posit configurations with lower number of fraction bits have lower power consumption but higher error.

In this work, we prioritise error over other metrics and hence select fixed-posits (32, 6, 2) and (32, 7, 1) that have the least error. Amongst them, (32, 6, 2) is the better representation because it's multiplier has lower power, area and delay. The results show that the fixed-posit (32, 6, 2) multiplier has 47\% lower power, 35.6\% lower delay and 38.5\% area savings compared to a (32, 6) posit, with zero relative error. 

We observe a similar trend for other bit-widths as well. However, we do not present results for all bit-widths due to space constraints. Even for OpenBLAS and AxBench benchmark, we observe that (N,6) posit multiplier and (N,6,2) fixed-posit multiplier have the same output error. However, owing to simpler design, fixed-posit multipliers have lower power. Overall, the results show that by fixing the bit format of posits, one can gain significant benefits in power, delay and area consumption with minimal error.

\begin{table}[]
\caption{Comparison of Area and Delay}
\label{tab:area-delay}
\resizebox{1\linewidth}{!}{%
\begin{tabular}{ccccccccc}
\hline
\multicolumn{1}{|c|}{\textbf{\begin{tabular}[c]{@{}c@{}}Bitwidth (N)\\ in (N, 6, 2)\end{tabular}}} & 32   & 30   & 28   & 26   & 24   & 22   & 20   & \multicolumn{1}{c|}{18}   \\ \hline
\multicolumn{1}{|c|}{\textbf{Relative Area}}                                                       & 1.09 & 0.94 & 0.84 & 0.72 & 0.63 & 0.56 & 0.39 & \multicolumn{1}{c|}{0.34} \\ \hline
\multicolumn{1}{|c|}{\textbf{Relative Delay}}                                                      & 1.04 & 1.00 & 0.98 & 0.93 & 0.89 & 0.85 & 0.80 & \multicolumn{1}{c|}{0.74} \\ \hline
\multicolumn{9}{c}{Area and Delay are relative to 32-bit IEEE-754}                                                                                                              \\ \hline
\end{tabular}
}
\vspace{-0.5cm}
\end{table}

\vspace{-0.5cm}
\subsection{Comparison with IEEE-754 Floating Point}
We compare (N, 6, 2) fixed-posit multipliers with 32-bit IEEE-754 floating point multiplier for varying values of bit-width (N). For a fixed-posit with bit-width N, we choose  (N, 6, 2) since it has lower area, power and delay among fixed-posits having least error.

\subsubsection{Workloads Used}
For evaluation, we use workloads from OpenBLAS\cite{openblas} and AxBench benchmarks\cite{axbench}. OpenBLAS benchmark consists of linear algebra workloads like matrix multiplications and eigen value computation. We observe that on average 50\% of floating point instructions executed by OpenBLAS workloads perform multiplication, making them ideal for evaluation. AxBench is an approximate computing benchmark and consists of error resilient applications like blackscholes, fft, kmeans, jpeg encoder and sobel. Further, we also evaluate our fixed-posit multiplier on two machine learning workloads. Our first workload performs inference using a three-layer fully connected network (FCN) on MNIST\cite{mnist}. Our second workload performs inference using ResNet-18\cite{resnet} on 5000 images of ImageNet's \cite{imagenet} validation set. 

\subsubsection{Experimental Methodology}

We compare the multipliers based on power, area and delay. We also compute the quality loss in the final output of the program on replacing IEEE-754 multiplications with fixed-posit multiplications.

We design the fixed-posit multipliers in Verilog and synthesize them as per details in Section \ref{sec:compposit}. We use FloPoCo\cite{flopoco} to generate the design of 32-bit IEEE-754 multiplier that we use as our baseline. The generated design does not include hardware for dealing with subnormal numbers. We obtain the area and delay results by synthesizing the designs for maximum frequency (zero slack). We also obtain a technology dependent netlist from this synthesis.

To perform the error and power analysis, we develop a pintool using Intel Pin\cite{pin}. Our pintool replaces all IEEE-754 multiplications in a program with fixed-posit multiplications during the program's execution. It does so by reading the operands from the required registers or memory locations, emulating the fixed-posit multiplication in software and storing the result in the destination register. While doing so, it also logs the trace of input operands which is used for power analysis. We report the loss in quality metric of the workloads by comparing the final outputs of the workload obtained on using IEEE-754 and fixed-posit multiplications. 

We divide our experiments for obtaining power into two phases. In the first phase, we procure the trace of input operands of multiplication for the workload using the pintool. In the second phase we convert this trace into Switching Activity Interchange Format (SAIF) files by passing it through Synopsys VCS along with technology dependent netlist and a Verilog testbench. Finally, we obtain the power using Synopsys Design Compiler by incorporating the SAIF files at 1GHz frequency. For few workloads, the trace of input operands is greater than 1 GB in size, making their usage prohibitively expensive in terms of computing resources required. Hence, for such workloads we sample 10 chunks of 10K consecutive multiplications, uniformly at random from all multiplications which are executed by the workload. This way we use 100K multiplications for further analysis.

We build the OpenBLAS and AxBench benchmarks for x86\_64 architecture using gcc 7.5.0 for nehalem target. Specifying Nehalem target prevents the compiler from using AVX instructions which is essential since our pintool cannot replace multiplications performed by AVX instructions. For our experiments, we execute OpenBLAS workloads for input size of 200 and AxBench workloads with the test inputs provided in the benchmark. For running the machine learning experiments we modify the PyTorch framework to emulate fixed-posit multiplications in software.


\begin{figure*}[ht]
\centering
\begin{minipage}[b]{.47\textwidth}
  \includegraphics[width=1.1\linewidth]{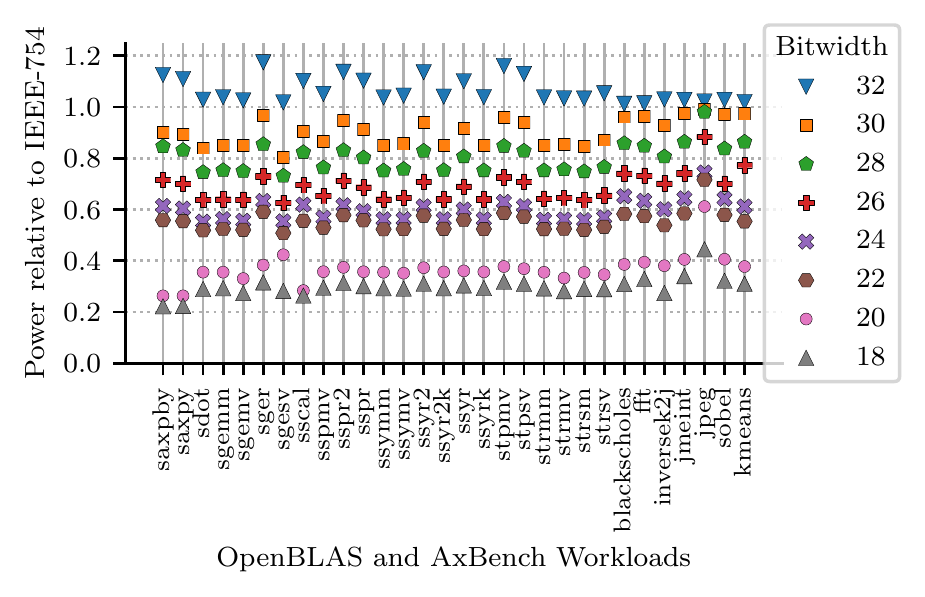}
  \caption{\centering Power Comparison on OpenBLAS (saxpby-strsv) and AxBench (blackscholes-kmeans) workloads}
  \label{fig:power}

\end{minipage}\qquad
\begin{minipage}[b]{.47\textwidth}
  \includegraphics[width=1.1\linewidth]{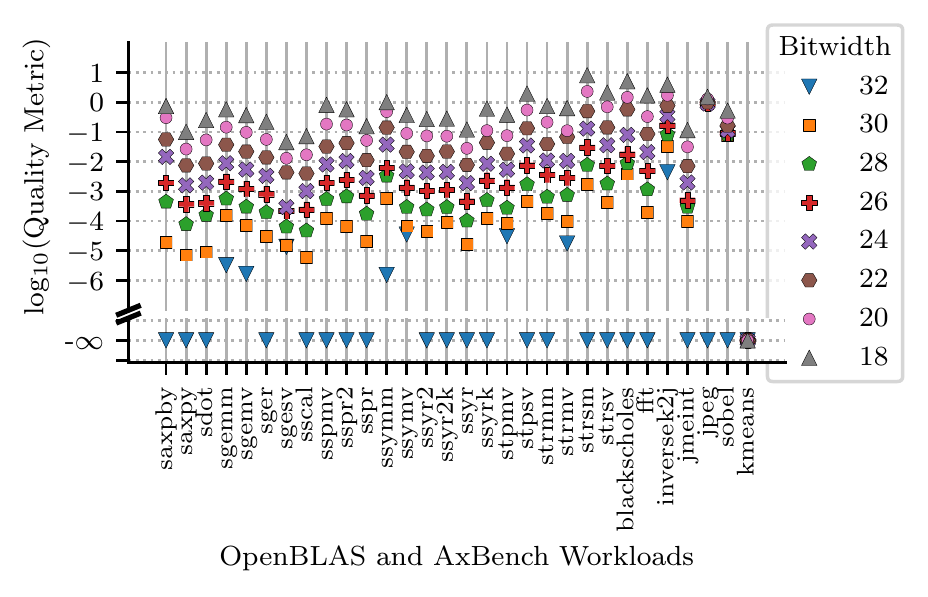}
  \caption{\centering Error Comparison on OpenBLAS (saxpby-strsv) and AxBench (blackscholes-kmeans) workloads.}
  \label{fig:error}
\end{minipage}
 \vspace{-0.5cm}
\end{figure*}

\subsubsection{Results}
\begin{table}[]
\centering
\caption{Accuracy drop (\%) for fixed-posits on ML applications compared to single precision IEEE-754 floating point}
\label{tab:acc-drops}
\begin{tabular}{@{}c|cccccccc@{}}
\toprule
\textbf{Bitwidth (N) in (N, 6, 2)} & \textbf{18} & \textbf{20} & \textbf{22} & \textbf{24} & \textbf{26} & \textbf{28} & \textbf{30} & \textbf{32} \\ \midrule
FCN on MNIST                      & 0.04        & 0.03        & 0           & 0           & 0           & 0           & 0           & 0           \\
ResNet-18 on ImageNet              & 0.12        & -           & -           & -           & -           & -           & -           & -           \\ \bottomrule
\end{tabular}
\vspace{-0.5cm}
\end{table}

Table \ref{tab:area-delay} shows the area consumption and circuit delay of fixed-posit multiplier implementations relative to 32-bit IEEE-754 multiplier. Both area consumption and circuit delay are independent of the workload. As we move from (32, 6, 2) to (18, 6, 2) fixed-posit representation, we observe that the area occupied by the multiplier and its delay decreases monotonically. 

Figure \ref{fig:power} shows the total power consumption (dynamic and leakage) of fixed-posit multipliers relative to 32-bit IEEE-754 multiplier for all workloads of OpenBLAS and AxBench. We observe that for all workloads the power consumption decreases as we reduce the bit-width of our fixed-posit representation. 

Figure \ref{fig:error} shows the quality metrics of the final output of a workload when we replace IEEE-754 floating point multiplications with (N,6,2) fixed-posit multiplications, on OpenBLAS and AxBench respectively. For OpenBLAS workloads the output quality is determined using relative error. Among AxBench workloads, the output quality metric for blackscholes, fft and inversek2j is relative error, for jmeint is miss rate, and for jpeg, sobel and kmeans is root mean squared error (RMSE). These quality metrics are specified by the authors of AxBench.

The output quality metric degrades as we go from (32, 6, 2) to (18, 6, 2) for all workloads. For majority of workloads, we see zero output quality loss when using (32,6,2) fixed-posit. Hence, Figure \ref{fig:error} shows the log of the quality loss to be negative infinity. Overall, the results show that the (32, 6, 2) fixed-posit multiplier is worse than 32-bit IEEE-754 floating point multiplier. The (32, 6, 2) multiplier occupies 9\% more area and also has 4\% longer delay. Further, it also consumes 6.3\% more power, on average, across all workloads, when compared to 32-bit IEEE-754 floating point multiplier. Finally, we observe a non-zero output quality loss for few workloads when using (32, 6, 2) fixed-posit. Hence, we only observe negative tradeoffs when replacing 32-bit IEEE-754 floating point with (32, 6, 2) fixed-posit.

We do start observing benefits when we reduce the bit-width of our fixed-posit. For (30, 6, 2) fixed-posit we observe 6\% area savings and 9.7\% power savings. Further, (30, 6, 2) has same delay as IEEE-754 floating point and has an average relative error of $1.6 \times 10^{-3}\%$ and average RMSE of 0.49. On the other extreme, we observe that (18,6,2) fixed-posit leads to 66\% area savings and 26\% reduction in circuit delay. We also observe 70.1\% reduction in power. However, we encounter an average relative error of $1.2\%$ and RMSE of 1.02. 

For image based workloads, jpeg and sobel, we also compute the peak signal to noise ratio (PSNR) averaged over 37 images. For (32, 6, 2) we obtain a PSNR of 100dB for both. For (18, 6, 2) the PSNR reduces to 45dB and 53.8dB for jpeg and sobel respectively. However, it is higher than visually acceptable quality threshold of 30dB used by prior work~\cite{approxlp}. 

Neural networks are yet another set of error-resilient applications, and we show the use of fixed-posit for these applications. Table \ref{tab:acc-drops} shows that on running a 3-layered FCN on MNIST with fixed-posit multiplications, there is zero drop in accuracy as we go from (32, 6, 2) to (22, 6, 2). For (18, 6, 2) we observe  0.04\% accuracy drop. For ResNet-18  on ImageNet, running experiments for each fixed-posit configuration takes significantly long time since ResNet-18 has $\sim$ 0.9 trillion multiplications. We report results for the worst case condition, i.e. (18, 6, 2) fixed-posit as it gives the highest quality loss for the rest of the applications. Interestingly, we observe only a 0.12\% accuracy drop, from 70.38\% to 70.26\%, on running ResNet-18 on ImageNet with (18, 6, 2) fixed-posit multiplications making a case for use of fixed-posits for error-resilient applications.

The results open up the possibility of large design space exploration for designers in the domain of approximate computing which is targeted for error-resilient applications. As different applications have different levels of error tolerance, the designer can trade-off quality for low power using fixed-posit. Fixed posit also  consumes lesser area compared with posit implementation, which is an additional advantage. 

\vspace{-0.2cm}
\section{Conclusion and Future work}
\vspace{-0.1cm}
In this work, we propose fixed-posit representation as a power-efficient alternative to conventional posits. Our evaluation shows that fixed-posit multipliers are area and power efficient than posit multipliers. We show that one can obtain significant improvements in power (up to  70\%), area (up to  66\%) and delay (up to  26\%) by using a fixed-posit multiplier over a 32-bit IEEE-754  floating  point  multiplier  with  minimal  relative output error (1.2\% average relative error). In this work, we only focus on multipliers. Performing similar analysis for an entire floating point unit which includes other operations will be crucial to conclusively observe the trade-offs of using fixed-posits over IEEE-754 floating point.

\vspace{-0.2cm}
\section*{Acknowledgment}
\vspace{-0.1cm}
We thank Dr. Farhad Merchant, RWTH Aachen University, for initial discussion about posits and Dr. Chandan Jha, IIT Bombay, for his help with machine learning experiments.
\end{document}